# Strain Engineering a $4a\times\sqrt{3}a$ Charge Density Wave Phase in Transition Metal Dichalcogenide 1T-VSe$_2$

Duming Zhang[1,2], Jeonghoon Ha[1,2], Hongwoo Baek[1,3], Yang-Hao Chan[4], Fabian D. Natterer[1,5], Alline F. Myers[1], Joshua D. Schumacher[1], William G. Cullen[1,2], Albert V. Davydov[6], Young Kuk[3], M. Y. Chou[4,7], Nikolai B. Zhitenev[1], and Joseph A. Stroscio[1]*

[1]Center for Nanoscale Science and Technology, National Institute of Standards and Technology, Gaithersburg, MD 20899, USA
[2]Maryland NanoCenter, University of Maryland, College Park, MD 20742, USA
[3]Department of Physics and Astronomy, Seoul National University, Seoul, 151-747, Korea
[4]Institute of Atomic and Molecular Sciences, Academia Sinica, Taipei 10617, Taiwan
[5]École Polytechnique Fédérale de Lausanne, CH-1015 Lausanne, Switzerland
[6]Material Measurement Laboratory, National Institute of Standards and Technology, Gaithersburg, Maryland 20899, USA
[7] School of Physics, Georgia Institute of Technology, Atlanta, Georgia 30332, USA

*Abstract*: We report a rectangular charge density wave (CDW) phase in strained 1T-VSe$_2$ thin films synthesized by molecular beam epitaxy on *c*-sapphire substrates. The observed CDW structure exhibits an unconventional rectangular $4a\times\sqrt{3}a$ periodicity, as opposed to the previously reported hexagonal $4a\times 4a$ structure in bulk crystals and exfoliated thin layered samples. Tunneling spectroscopy shows a strong modulation of the local density of states of the same $4a\times\sqrt{3}a$ CDW periodicity and an energy gap of $2\Delta_{\text{CDW}} = (9.1 \pm 0.1)$ meV. The CDW energy gap evolves into a full gap at temperatures below 500 mK, indicating a transition to an insulating phase at ultra-low temperatures. First-principles calculations confirm the stability of both $4a\times 4a$ and $4a\times\sqrt{3}a$ structures arising from soft modes in the phonon dispersion. The unconventional structure becomes preferred in the presence of strain, in agreement with experimental findings.

---

* Correspondence and requests for materials should be addressed to J.A.S. (joseph.stroscio@nist.gov)

## I. Introduction

Lattice deformations induced by strain alter the interatomic distances in crystalline solids and modify both the electronic levels of bonding electrons and phonon frequencies of the ionic lattice. For that reason, *strain engineering* has been used to tailor electronic properties of semiconductor nanocrystals [1], and has been applied to 2D materials such as graphene [2–6], and more recently to layered transition metal dichalcogenides (TMDs) [7–11]. The work on strain effects in TMDs has largely focused on band gap engineering, and resulting optical, magnetic, and conductivity responses [8–11]. Many TMD materials also support charge density wave (CDW) phases, whose structures and energetics will sensitively depend on strain. This follows from the influence of strain on the electron-phonon coupling. The present work explores strain engineering of a rectangular CDW phase in TMD 1T-VSe$_2$ thin films grown on sapphire substrates.

CDWs are broken symmetry states in metals brought about by electron-electron and electron-phonon interactions, resulting in a periodic spatial modulation of the charge density [12,13]. CDWs are among the most intriguing phenomena in condensed matter physics and remain a subject of intense research as they are encountered in many highly correlated materials and low-dimensional electron systems, such as cuprates [14,15] and TMDs [16–25]. A renewed interest in CDW in TMDs is fueled by the immense impact of heterostructure devices that contain a variety of van der Waals-bonded two-dimensional (2D) layered materials [26–29], among those, graphene and hexagonal boron nitride are the most prominent examples [30–32]. TMDs have a stoichiometry $MX_2$, where $M$ is a transition metal and $X$ is the chalcogen, Se, Te, or S. Ultra-thin sheets of layered TMDs are particularly interesting owing to their chemically and electronically versatility [33]. They can be incorporated in device heterostructures as superconductors, metals, semiconductors, or insulators, depending on the combination of the $M$ and $X$ elements, crystal structure, and layer thickness [33–35]. Recent research focused on the possibility to tune the CDW phase transition in TMDs through a gate potential and via layer thickness for electronic switches, logic, and other device applications [36–40]. An alternative route to modify the CDW structure is intercalation that can change the electronic structure and induce strain. 1T-TaS$_2$ is a case in point, as a new rectangular $c(2\sqrt{3}a \times 4a)$ structure is formed from the intercalation of alkali metals such as Na and Rb [41–45], similar to that found in the strained 1T-VSe$_2$ studied here.

1T-VSe$_2$ is one of the layered TMDs that exhibits a phase transition to a CDW state at approximately 100 K [46]. It is one of the few examples of an exotic $4a \times 4a \times 3c$ three-dimensional CDW with a large component perpendicular to the 2D layers [47–49]. Recent transport measurements have shown a significant dependence of the CDW transition temperature upon layer thickness [36]. In addition, ultra-thin VSe$_2$ nanosheets were found to be ferromagnetic at room temperature [50] and exfoliated samples show an interesting 3D to 2D crossover [51]. The initial explanation for CDW formation was introduced by Peierls. He conjectured that a one dimensional metal is unstable to the formation of a CDW with a periodicity related to the Fermi surface spanning vectors $2k_F$ [52]. More recent theory has focused on the role of momentum-dependent electron-phonon coupling *versus*

Fermi surface nesting in examining the origins of CDW formation in various layered materials [20,25].

In this article, we report on *in situ* scanning tunneling microscopy/spectroscopy (STM/S) measurements on strained epitaxial 1T-VSe$_2$ thin films synthesized by molecular beam epitaxy (MBE) on *c*-sapphire substrates without exposure to atmosphere. STM measurements show a previously unobserved $4a\times\sqrt{3}a$ rectangular CDW structure in the VSe$_2$ thin films at temperatures of 4 K and below. Tunneling spectroscopic measurements show a soft CDW energy gap of approximately $2\Delta_{\text{CDW}} = (9.1 \pm 0.1)$ meV [53]. At temperatures below 500 mK, a full gap opens up in the tunneling spectrum showing a transition to a fully insulating state. First-principles calculations identify this additional CDW state to be consistent with soft modes in the normal-state phonon dispersion and as a competing low energy state to the conventional $4a\times 4a$ CDW. We note that a conventional $4a\times 4a$ CDW was observed in thin (10 nm) exfoliated VSe$_2$ layered samples [51]. This indicates the strain in our epitaxial film plays an important role in the rise of the unconventional $4a\times\sqrt{3}a$ CDW. The organization of this manuscript is as follows. In section II we describe the molecular beam epitaxial growth of strained VSe$_2$ thin films on *c*-sapphire, and the characterization by TEM and X-ray diffraction to determine the epitaxial relationship. In Section III we describe the scanning tunneling microscopy and spectroscopy results, which display the rectangular $4a\times\sqrt{3}a$ CDW structure in the VSe$_2$ thin films, and the corresponding energy gaps. In the Section IV we show first-principles calculations of the phonon dispersion for pristine and strained VSe$_2$, and compare structural models with experimental STM data. Finally, we conclude in Section V.

## II. Molecular Beam Epitaxial Growth of Strained VSe$_2$ Thin Films

TMDs do not have the same inertness as graphene with respect to exposure to atmospheric conditions [54]. For device applications employing few layers or single layer TMDs, a major challenge is to keep ultra-thin films clean from an exposure to atmosphere for *in situ* measurements. TMDs, including VSe$_2$, are often synthesized by chemical vapor transport methods yielding bulk single crystals [46,47,55–59]. Only one study, to our knowledge, reports on the preparation of VSe$_2$ by molecular beam epitaxy in growing (SnSe)(VSe$_2$) misfit layered compounds [60]. The advantages of MBE growth are numerous: composition can be adjusted by tailoring flux ratios, films thickness can be controlled down to monolayer, strain can be induced with suitable substrate lattice mismatch or due to difference in thermal expansion between the substrate and TMD film, and post growth measurements can be performed *in situ* without exposing the film to atmosphere.

The VSe$_2$ thin films were MBE grown on *c*-sapphire using high purity vanadium (99.9 %) and ultra-high purity selenium (99.999 %). The growth temperature was ≈ 775 ºC and the growth rate was ≈ 0.75 nm/min with a Se/V flux ratio of 12.5. The sapphire substrates were cleaned in isopropyl alcohol/acetone and annealed in an oxygen atmosphere at 950 ºC for one hour prior to the growth to improve their (0001) surface morphology. The as-grown samples were transferred in ultra-high vacuum into a cryogenic STM system operating at temperatures down to 10 mK [61]. Sample temperatures almost coincide with the dilution refrigerator temperatures down to 150 mK. At lower temperatures, the sample temperatures remain at (150±15) mK due to RF heating determined from fitting of superconducting Al spectra [53,62]. The STM system contains facilities for MBE growth of samples and probe tip preparation. Probe tips were fabricated from Ir wire by electro-chemically polishing and further cleaning was done by electric field evaporation in a field-ion-microscope. Tunneling spectroscopy was performed by measuring the tunneling differential conductance, *dI/dV*, using lock-in amplifier detection with modulation frequency of 141 Hz and a AC modulation voltage added to the DC sample bias.

1T-VSe$_2$ has a hexagonal layered crystal structure with the $P\bar{3}m1$ space group. Each VSe$_2$ layer consists of three Se-V-Se atomic layers arranged in an octahedral coordination and is weakly bonded to the next trilayer by the van der Waals force, as shown in Fig. 1(a). As expected for van der Waals layered materials, VSe$_2$ thin films grow layer by layer, as confirmed by reflection high-energy electron diffraction (RHEED) oscillations observed during the MBE growth (Fig. 1d). X-ray diffraction measurements shown in Fig. 1(e) verify the 1T crystal structure with a slightly smaller lattice constant along the *c*-axis, $c_{film}$ = (0.590 ± 0.001) nm [53] compared to bulk crystals, $c_{bulk}$ = 0.611 nm [46,63,64]. Cross-sectional high-resolution transmission electron microscopy (HRTEM) measurements of a specimen with a thickness of 80 layers (≈ 47 nm) are shown in Figure 1(b). The clear phase contrast indicates the single-crystalline nature of the VSe$_2$ film in registry with the substrate. The selective area diffraction pattern in Fig. 1(c) reveals the following epitaxial relationship between the film and the substrate: $(1\bar{1}00)_{sapphire}$ || $(1\bar{2}10)_{VSe2}$ and $(0001)_{sapphire}$ || $(0001)_{VSe2}$. The corresponding schematic atomic epitaxial model is shown in Fig. 1(f).

The slightly smaller lattice constant along the *c*-axis, determined by X-ray diffraction, was also seen in HRTEM, along with a corresponding expansion of the in-plane lattice constant of $a_\text{film} = (0.349 \pm 0.002)$ nm [53] compared to the bulk value of $a_\text{bulk} = 0.335$ nm [46]. Both the compression along the *c*-axis and expansion along the *a*-axis indicate the presence of strain that modifies the relative energetics of possible CDW states, as discussed in Section IV. The strained VSe₂ layer growth on sapphire substrates can be understood by comparing their coefficients of thermal expansion (CTE) in relation to the cooling process after the MBE growth. Notably, the CTE of VSe₂, $\alpha_a = 35\times10^{-6}$ K$^{-1}$ [65] is approximately four times larger than that of sapphire, $\alpha_a = 8.6\times10^{-6}$ K$^{-1}$ [66]. During the post-growth temperature drop of ≈700 ºC, the sapphire substrate prevents the VSe₂ layer from shrinking, which causes in-plane biaxial tensile stress in the film and explains a larger $a_\text{film}$ lattice parameter relative to $a_\text{bulk}$. Consequently, the in-plane tensile strain is compensated by compressive strain in the orthogonal direction through the Poisson effect thus causing a reduction of the *c* lattice parameter in the VSe₂ film, as confirmed by XRD and HRTEM measurements.

The morphology of the thin film was examined by *in situ* STM measurements. A topographic image in Fig. 1(g), from a large area of a thin film sample shows islands with the hexagonal symmetry of the VSe₂ lattice. The orientation of the VSe₂ islands edges suggests a preferred in-plane orientation with respect to the substrate despite small misalignment among different islands. This is in agreement with the epitaxial relationship between the VSe₂ and sapphire substrate discussed above. We also note that the VSe₂ island edges are typically aligned at 60º and 120º from each other. This three-fold symmetry is also observed in images of grain boundaries [Fig. 1(h)], which are typically oriented at angles of 120º. However, inspection of the CDW structure on the grain boundary terraces shows a 2-fold symmetry, instead of the expected 3-fold symmetry found in bulk crystals, indicating a new CDW structure in these VSe$_2$ strained films.

## III. The $4a\times\sqrt{3}a$ CDW Structure in VSe$_2$

Figure 2 shows the structure and symmetry of the strained VSe$_2$ $4a\times\sqrt{3}a$ CDW phase. This is strikingly different from the previously observed $4a\times 4a$ commensurate CDW that followed the hexagonal symmetry of the underlying VSe$_2$ lattice, and had a transition temperature of ≈100 K [36]. The atomic structure of the CDW is apparent with decreasing tip-sample separation in the STM measurements shown in Figs. 2(a)-2(c). The unit cell is outlined in the STM image in Fig. 2(c) and in the FFT image in Fig. 2(d). The small number of presumably Se vacancies at the surface emphasizes the quality of the film. The line profiles of the CDW shown in Figs. 2(e) and (f) correspond to the white lines in Fig. 2(c). Along the $<11\bar{2}0>$ direction, we observe the conventional CDW $4a$ periodicity, as shown in the height trace in Fig. 2(e). However, along the $<1\bar{1}00>$ direction, the CDW has an unconventional $\sqrt{3}a$ periodicity, as observed in the height trace in Fig. 2(f). This CDW pattern is very robust under various imaging conditions, and was observed in all locations surveyed in each sample, and in five different samples of varying thickness. In addition, the CDW structure remained intact up to 4 K, the highest temperature achievable in our ultra-low temperature STM system.

The CDW wavelength is constant over a wide energy range [Fig. 3] indicating the CDW is associated with an ionic change in the lattice structure. However, the intensity and phase of the CDW varies greatly with energy, particularly in comparing filled vs empty state images in Fig. 3. The lattice distortion from a CDW is usually accompanied by corresponding gaps at the Fermi surface. The strong phase change across the Fermi level is indicative of a CDW energy gap [67–70]. At temperatures below the CDW transition temperature, $T_{CDW}$, this gap can be measured as a gap ($2\Delta$) in tunneling spectroscopy. The VSe$_2$ CDW gap has only been reported in a few studies by STM measurements. Early pioneering measurements reported an energy gap of ≈40 meV at 4.2 K [71], while more recent investigations found a gap of ≈80 meV at 60 K [72]. These reported gaps were not full gaps with zero conductance, but rather soft gaps defined by local maxima nearest to the Fermi level. Both these values lead to very large ratios of $2\Delta/k_B T_{CDW}$, ≈ 4.7 to ≈ 9.3, with a transition temperature of $T_{CDW}$ ≈100 K. These large ratios may indicate a possible incorrect assignment of spectral features with a CDW energy gap in these early measurements. Additionally, as shown below, a large amount of state density is modulated at the CDW period within this large energy range, hence we think these early results could have had an incorrect assignment of the CDW gap energy.

We obtain the local density of states (LDOS) of the $4a \times \sqrt{3}a$ CDW phase by measuring the bias and spatial dependent $dI/dV$ spectra. Figures 4(a) and 4(b) show tunneling spectra obtained in two different locations along the $<11\bar{2}0>$ axes. The spectra are characterized by four features: 1) Rather weak peaks where strong coherence peaks would be expected for a fully gaped CDW [see Figs. 4(a) and 4(b)], 2) a "V" shaped gap, 3) a small zero conductance gap straddling the Fermi level at zero bias, and 4) a strong modulation of the peak at ≈+10 mV depending on spatial position, as seen in comparing the intensity at 10 mV in Figs. 4(a) and 4(b). Figures 4(d) and 4(e) show a series of tunneling spectra as a function of distance along the two crystallographic directions indicated in Fig. 4(c). In Fig. 4(d), the modulation at positive bias between the spectra in Fig. 4(a) and (b) is evident. Along both directions in Fig. 4(c), the spectra show a strong modulation of the LDOS with the respective CDW periodicity of $4a$ along the $<11\bar{2}0>$ direction [Fig. 4(d)] and of $\sqrt{3}\,a$ along the $<1\bar{1}00>$ direction [Fig. 4(e)]. We associate the "V" shaped gap in the spectra with a soft CDW gap indicating a partially gapped Fermi surface. In analogy with broadened superconducting spectra, a more direct measure of the gap are the inflection points inside of the gap, rather than the positions of the broadened coherence peaks. The latter shift to higher energies and thus overestimates the energy gap, as seen, for example, in temperature broadened spectra [62]. We consequently use the inflection points of the "V" shaped gap, determined by minima and maxima in the derivative of the differential conductance spectra, to systematically determine the CDW $2\Delta$ gap, as shown in Figs. 4(a) and 4(b) by the orange lines. The two sets of inflection points reveal two gaps. The smaller gap $\Delta_I$, indicated by the inner most inflection points, is a temperature dependent insulating gap, as discussed below. We associate the CDW gap with the outer most gap observed in Figs. 4(a) and 4(b). We determine a CDW gap averaged over multiple samples with a width of $2\Delta_{CDW} = (9.1 \pm 0.1)$ meV [53]. The comparatively small value of the CDW gap may have been smeared out by the higher temperature of previous measurements, or it may be a characteristic of this new strained layer CDW phase. Collectively, the strong phase contrast observed across the Fermi level in Fig. 3 [see also Fig. 6(f) and (g)], the LDOS periodic modulation, and the gap structure observed in Fig. 4, are strong evidence for a CDW phase in these strained layer VSe$_2$ films.

Interestingly, our VSe$_2$ thin films show a transition to an insulating phase upon cooling to below ≈500 mK. At temperatures below 500 mK the tunneling spectra are characterized by a full gap with zero conductance, which further opens at progressively lower temperatures. The size of the full gap is sample dependent, ranging from ≈2meV [see zero conductance region in Fig. 4(a) and (b)] to as large as ≈300 meV observed in one sample. All samples show the same CDW gap above 500 mK, and the onset of the insulating gap is only observed below this temperature. The measured gap width as a function of temperature is shown in Fig. 5(a), for one of the samples with an insulating gap of ≈30 meV at the lowest temperature of 150 mK. Above 500 mK the spectra show the characteristic "V" shaped CDW gap of ≈9 meV. Below 500 mK the zero bias conductance gradually drops to zero at ≈240 mK and then the full gap develops with decreasing temperature. Figure 5(b) shows the temperature dependent energy gap determined by the inflection points from the derivative spectra as described in Fig. 4. The energy gap is the largest (≈30 meV) at the lowest temperature and then decreases linearly with $1/T$, followed by a plateau at temperatures above 500 mK with a constant CDW gap. We estimate an energy scale from a linear fit of the data between 150 mK and 500 mK as (303 ± 14) μeV [53]. A similar evolution of a CDW to an insulator state was observed in thin films of $TaS_2$ and associated with entering a Mott insulating phase at low temperatures [38,73]. Likewise, the low temperature insulating phase observed here in strained layer $VSe_2$ films may be of a similar origin and possibly depends on the disorder in the thin films. This would explain the varying gap size for different samples.

### IV. Theory and Energetics of CDW States in $VSe_2$

Insight into the origin of the unconventional rectangular CDW on the strained $VSe_2$ thin films is obtained from first-principles calculations. The calculations were performed using the Vienna *ab initio* simulation package (VASP) [74–77] with the projector augmented wave (PAW) method [78,79]. For the exchange-correlation functional, we mainly used the generalized gradient approximation (GGA) of the Perdew-Burke-Ernzerhof (PBE) type [80] and compared our results to the local-density approximation (LDA). The theoretical in-plane lattice constant $a$ obtained by the PBE calculations was 0.335 nm, in excellent agreement with the bulk crystal lattice constant. However, the out-of-plane lattice constant $c$ was overestimated by 12 % compared with the bulk value, which is a known issue of the PBE functional for van der Waals materials. Therefore, we have fixed the $c$ lattice constant to the bulk value of 0.610 nm for the unstrained material in the calculation. The phonon calculations use density functional perturbation theory (DFPT) implemented in the Quantum Espresso package [77,81] with the LDA and a norm-conserving pseudopotential. The plane wave energy cutoff is 40 Hartrees, and a 16×16×10 $k$-mesh is applied. The phonon dispersions for the undistorted state are interpreted from results on a 4 × 4 × 2 $q$-mesh. The binding properties of the VSe$_2$ bulk and few layers has been investigated in a previous calculation [82], but the CDW energetics study has not been reported.

The 3D Brillouin zone is shown in Fig. 6(a). We first examine the phonon dispersions of bulk VSe$_2$ in the ideal state and in the strained state using the lattice parameters found experimentally in our thin films, as shown in Figs. 6(b) and 6(c), respectively. As expected, soft modes (negative frequencies) occur for both sets of lattice parameters, suggesting a CDW instability. Several instability groups in the bulk are shown in Fig. 6(b). One is located midway between $k$-points $A$ and $L$, reappearing also at midway between $\Gamma$ and $M$, as marked by red arrows. We have varied the wave vector along the $z$ direction and found that the instability actually peaks at $k_z \approx 2/3$ (in units of $2\pi/c$). This corresponds to a wave vector of $Q_1 = (1/4, 0, 2/3)$ (in units of reciprocal lattice vectors).

This wave vector and the equivalent ones, related by symmetry, suggest a CDW periodicity of $4a \times 1a \times 3c$, $1a \times 4a \times 3c$, or $4a \times 4a \times 3c$, in agreement with reports for bulk VSe$_2$ [83,84], and unstrained thin layered samples [51]. We have checked the phonon frequencies at the $\Gamma$ point for these phases and find that they are non-imaginary.

In addition, the phonon dispersion of the unstrained phase also exhibits other groups of soft modes along $\Gamma$-$K$ and $A$-$H$ in Fig. 6(b). Possible origins of these soft modes include, among others, strong electron-phonon interaction and Fermi surface nesting. Our calculated bare electronic susceptibility indicates that $Q_1$ is associated with a nesting feature. Further noticeable Fermi nesting appears at $Q_2$, around 1/3 of $AH$ (as indicated by a black arrow), which is compatible with a rectangular $4a \times \sqrt{3}a$ unit cell in the $xy$ plane in the sense that $Q_2$ is one of its reciprocal lattice vectors. By comparison, the phonon dispersion for the strained crystal exhibits more soft modes in Fig. 6(c), reflecting an enhanced trend toward lattice instability. It is therefore not surprising that the strained films resulting from the interaction with the substrate may generate different CDW patterns than the ideal bulk.

We next examine the energies of the various CDW states. Since the interlayer interaction is much weaker than the intralayer interaction for the layered compound VSe$_2$, we focus on the CDW energetics in two dimensions and optimize the atomic configurations in the $4a \times \sqrt{3}a$ and $4a \times 4a$ supercells of a single layer; energy results for the ideal and strained lattice parameters are shown in Table 1. At $a = 0.335$ nm the $4a \times 4a$ CDW phase is favored over the $4a \times \sqrt{3}a$ phase. However, in presence of strain ($a = 0.350$ nm) we find that both CDW phases are preferred over the normal phase and that the respective energy difference between the $4a\times4a$ and $4a\times\sqrt{3}a$ CDW phases becomes significantly smaller. This again indicates that the presence of substrate interactions may stabilize the $4a\times\sqrt{3}a$ CDW phase, as revealed by the present STM study.

For the strained lattice, we find one metastable $4a\times\sqrt{3}a$ CDW phase that closely matches the experimental images. (We have checked the phonon frequencies at Γ of this structure and did not find imaginary frequencies.) Figure 6(d) shows the top view of the atomic structure for this $4a\times\sqrt{3}a$ CDW phase determined from theory. For comparison, Figs. 6(f) and 6(g) display the experimental STM images of the CDW structure for filled and empty electronic states viz. on either side of the CDW gap, respectively. Comparing the structural model Fig. 6(d) with the filled state image Fig. 6(f), we observe that the minima in the four corners of the outlined unit cell arise from Se atoms, which are strongly bonded to only one V atom, versus the normal bonding to two V atoms. A corresponding simulated STM image of this structure [Fig. 6(e)] shows reasonable agreement with the filled state experimental STM image in Fig. 6(f), and suggests that strain plays a major role in stabilizing the rectangular $4a\times\sqrt{3}a$ CDW phase in MBE grown VSe$_2$ thin films on c-sapphire substrates.

## V. Conclusions

In conclusion, we have synthesized a strained crystal structure of VSe$_2$ thin films on sapphire substrates by MBE. A new rectangular CDW structure is observed at low temperatures with $4a\times\sqrt{3}a$ periodicity. First-principles calculations indicate that this CDW phase is consistent with soft phonon modes already present in the bulk crystal, which become more energetically favorable in the presence of strain. Tunneling spectroscopy shows a strong modulation with the CDW lattice periodicity and determines a CDW energy gap, $2\Delta_{CDW} = (9.1 \pm 0.1)$ meV, consistent with the strong phase contrast observed in the bias dependent STM images. The CDW transforms to an insulating low temperature phase below 500 mK, as indicated by the opening of a full insulating gap in spectroscopic measurements. For future studies, these results demonstrate the potential of strain engineering as a promising avenue to explore new emerging phases in low dimensional materials and in particular for cases where the sensitive balance of competing interactions can be manipulated and tuned.

**Acknowledgements**
D.Z., J.H., and W.G.C. acknowledge support under the Cooperative Research Agreement between the University of Maryland and the National Institute of Standards and Technology Center for Nanoscale Science and Technology, Grant No. 70NANB10H193. H.B. and Y.K. are partly supported by Korea Research Foundation through Grant No. KRF-2010-00349. F.D.N. appreciates support from the Swiss National Science Foundation under project number PZ00P2_167965. M.Y.C. acknowledges the support by the U.S. National Science Foundation (EFMA-1542747). Y.-H.C. is supported by a thematic project at Academia Sinica.

Table 1 Calculated charge density wave energies for different supercells in monolayer VSe$_2$. The values are determined in meV per chemical unit with respect to the energy of the normal state. *Metastable state with a structure closely resembling the experimentally observed phase (see Fig. 6).

| Monolayer VSe$_2$ | $4a \times \sqrt{3}a$ | $4a \times 4a$ |
|---|---|---|
| $a = 0.335$ nm | -0.2 | -8.4 |
| $a = 0.350$ nm | -31.3, -14.8* | -36.5 |

**Figure Captions**

FIG. 1. Molecular beam epitaxy growth of strained VSe$_2$ thin films with rectangular CDW structure. (a) Structure of 1T-VSe$_2$ consisting of trilayer elements of Se (red) and V (green). The upper layer Se are larger in the top view. (b) HRTEM image of the cross section of the MBE grown VSe$_2$ thin film. (c) Selective area diffraction pattern from the film shown in (b), which defines the following epitaxial relationship between the VSe$_2$ film and sapphire substrate: (0001) sapphire ∥ (0001) VSe$_2$ and ($1\bar{1}00$) sapphire ∥ ($1\bar{2}10$) VSe$_2$. (d) RHEED intensity oscillations observed during growth. (e) X-ray diffraction pattern of a ten-layer film of 1T-VSe$_2$ on *c*-sapphire. VSe$_2$ crystallizes in hexagonal structure [space group $P\bar{3}m1$ (164)] with a calculated *c*-lattice parameter of 0.590±0.001 nm [53]. (f) An epitaxial model of VSe$_2$ on *c*-sapphire derived from TEM showing 30º rotation for the VSe$_2$ lattice relative to the Al$_2$O$_3$ lattice. Only O (light green) and Se (dark green) atoms are shown. (g) STM image of VSe$_2$ islands grown on sapphire by MBE. (h) STM image of a grain boundary imaged in one of the VSe$_2$ islands. The STM images in (g,h) are shown in a color scale from dark to light, covering a height range of 24 nm and 0.1 nm, respectively.

FIG. 2. Rectangular CDW in MBE grown strained VSe$_2$ thin films. (a-c) STM topographic images, 14 nm × 14 nm, with decreasing tunneling impedance (decreasing sample-tip distance). Tunneling impedance: (a) 25 MΩ, (b) 12.5 MΩ, and (c) 6.25 MΩ. Height range for images: (a) 114 pm, (b) 117 pm, and (c) 107 pm. Tunneling current 4 nA. $T = 150$ mK. The white box in (c) outlines the $4a\times\sqrt{3}a$ unit cell. (d) 2D FFT of the image in (c) showing the $4a\times\sqrt{3}a$ unit cell. (e,f) STM height profiles from (c) along the indicated white lines showing the $4a$ periodicity along the $<11\bar{2}0>$ direction in (e) and the $\sqrt{3}a$ periodicity along the $<1\bar{1}00>$ in (f).

FIG. 3. Voltage dependent STM topographic imaging of the VSe$_2$ CDW. (a-c) Negative sample bias, -25 mV, -60 mV, and -200 mV, respectively. (d-f) Positive sample bias, 5 mV, 60 mV, and 150 mV, respectively. Image size 14 nm × 14 nm. The purple rectangle outlines the same position in all images. Tunneling impedances: (a) 6.25 MΩ, (b) 100 MΩ, (c) 333.3 MΩ, (d) 8.3 MΩ, (e) 100 MΩ, (f) 250 MΩ. Height range for images: (a) 107 pm, (b) 148 pm, (c) 113 pm, (d) 259 pm, (e) 146 pm, and (f) 121 pm. $T = 150$ mK.

FIG. 4. Tunneling spectroscopy of the rectangular CDW state in VSe$_2$ thin films. (a,b) *dI/dV* spectroscopy measurements (red lines) obtained on the lattice row with 4a periodicity, at the locations indicated by the dots in (c). Two energy gaps, $\Delta_{CDW}$ and $\Delta_I$, discussed in the text are defined from the inflection points in the *dI/dV* measurements, corresponding to the maxima and minima of the derivative of the differential conductance (orange lines). (c) STM image, 6 nm x 3 nm, at the locations of *dI/dV* spectroscopy measurements showing strong modulations in the local density of states of (d) $4a$ periodicity along the $<11\bar{2}0>$ direction and (e) $\sqrt{3}a$ periodicity along the $<1\bar{1}00>$ direction. A CDW gap of $2\Delta_{CDW} = (9.1 \pm 0.1)$ meV is determined from the spectra as indicated in (a) and (b) [53].

FIG. 5. Tunneling spectroscopy of the low temperature VSe$_2$ insulating state. (a) *dI/dV* spectra (symbols) as a function of temperature. The solid lines are smooth splines through the data points. (b) The 2Δ energy gap measured from the spectra in (a) as a function of $1/T$. At temperatures above 500 mK the energy gap remains constant with $2\Delta = (9.1 \pm 0.1)$ meV [53]. Uncertainties in the energy values were obtained from non-linear peak fitting of the derivative spectra of (a).

FIG. 6 Theory and experiment of a new CDW phase in strained VSe$_2$. (a) Fermi surface and 3D Brillouin zone for VSe$_2$. Phonon dispersions of the normal state of the (b) ideal and (c) strained bulk crystal of VSe$_2$. Imaginary frequencies corresponding to soft modes are shown as negative values in the plots. The red and black arrows indicate soft modes corresponding to the $4a\times4a$ and $4a\times\sqrt{3}a$ structures, respectively. (d) Top view of the atomic structure of the $4a\times\sqrt{3}a$ CDW phase in monolayer VSe$_2$ which matches closely with the STM results [see Table 1]. The red spheres represent Se atoms, (top layer larger spheres), and the green spheres the V atoms. The V-Se pairs with a contracted bond length are connected in the figure. The dashed lines outline the $4a\times\sqrt{3}a$ unit cell. (e) Simulated STM image for this structure at -25 mV bias at the isosurface density of $5\times10^{-5}\ e/a_0^3$. An experimental STM image of (f) filled states at a bias of –25 mV, and (g) empty states at a bias of 5 mV. The color scale covers a range of 66 pm and 80 pm, from dark to bright, in (f) and (g). The white rectangles in (e-g) outline the same $4a\times\sqrt{3}a$ unit cell areas. $T$ = 150 mK.

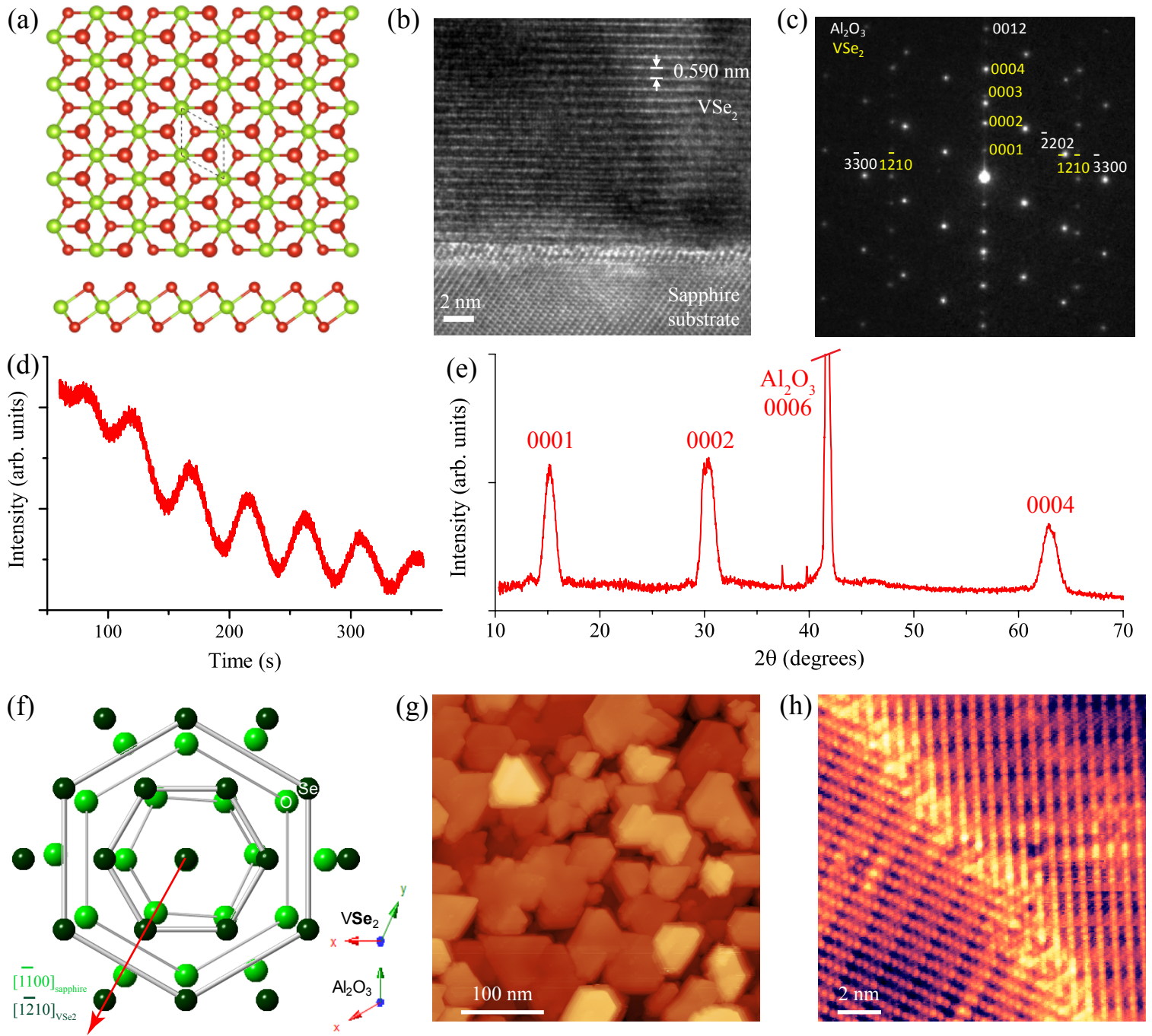

FIGURE 1.

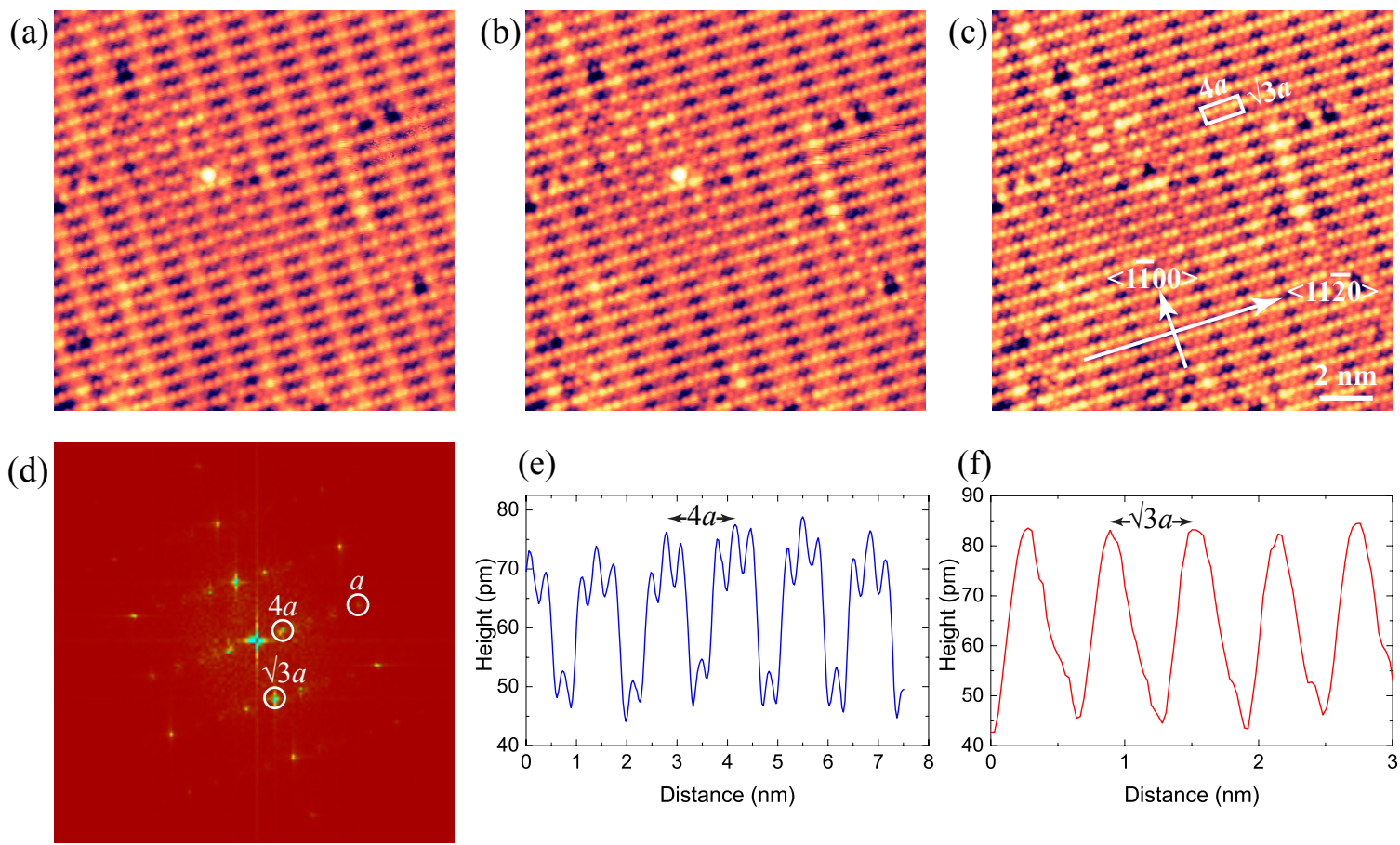

FIGURE 2.

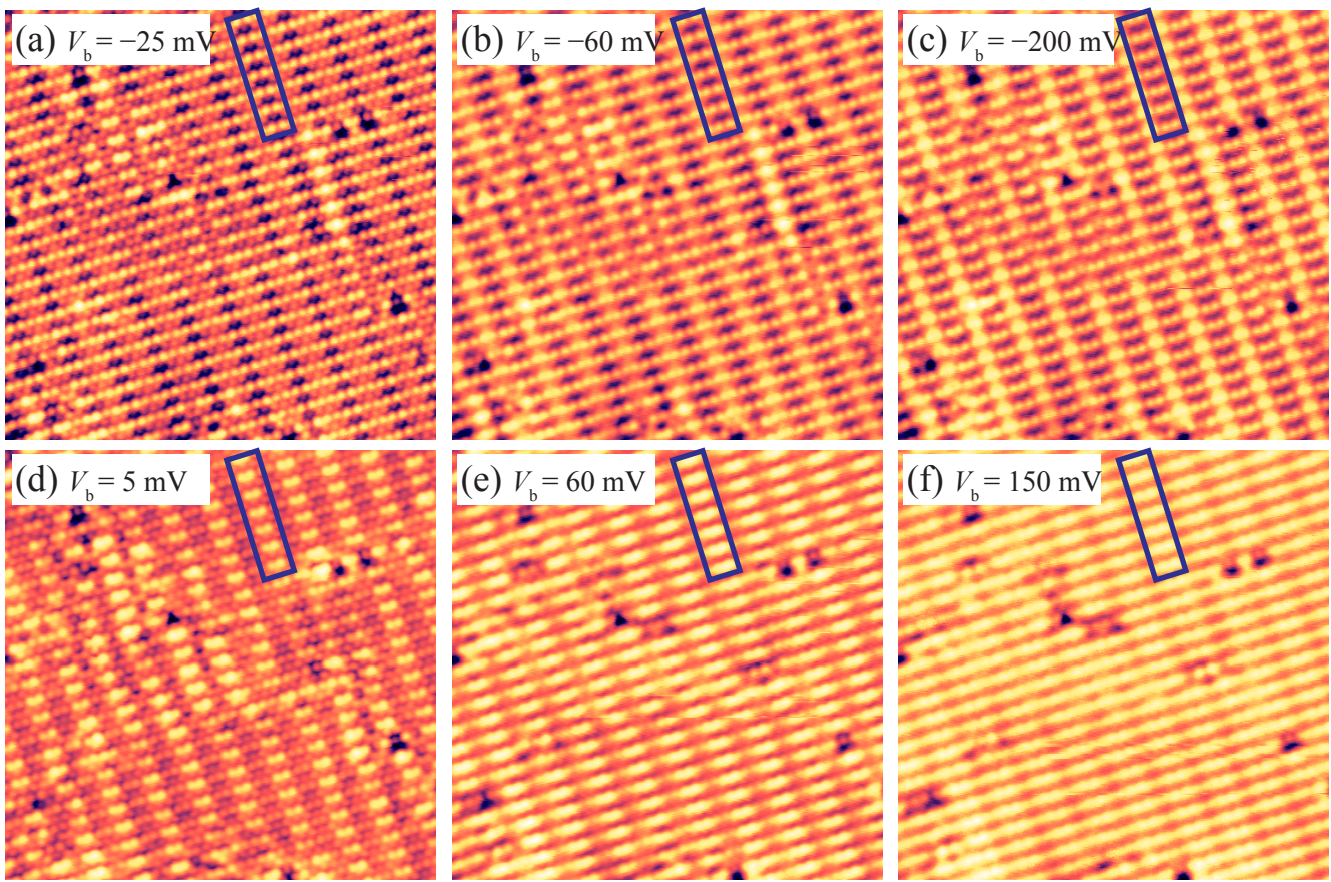

FIGURE 3.

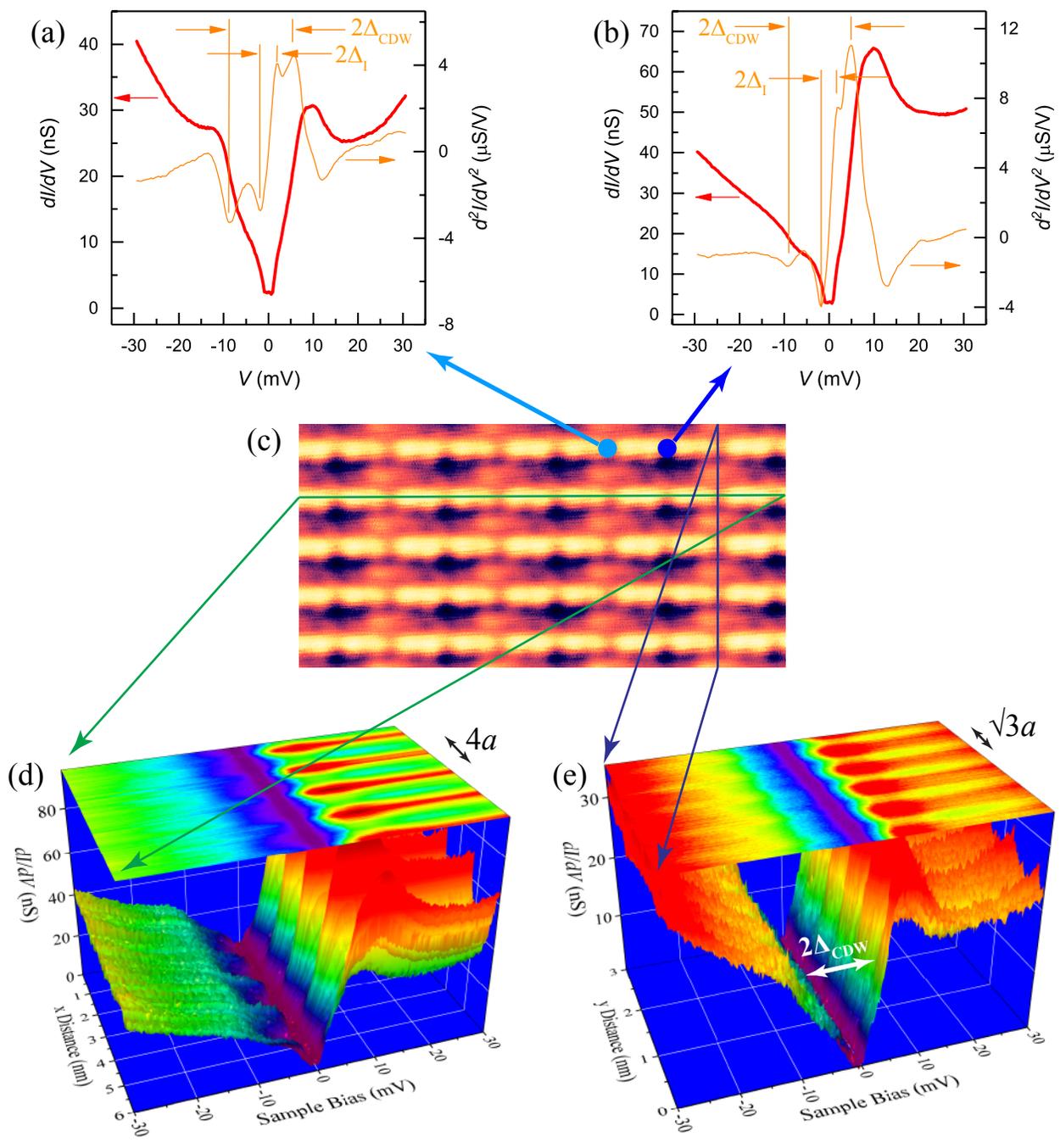

FIGURE 4.

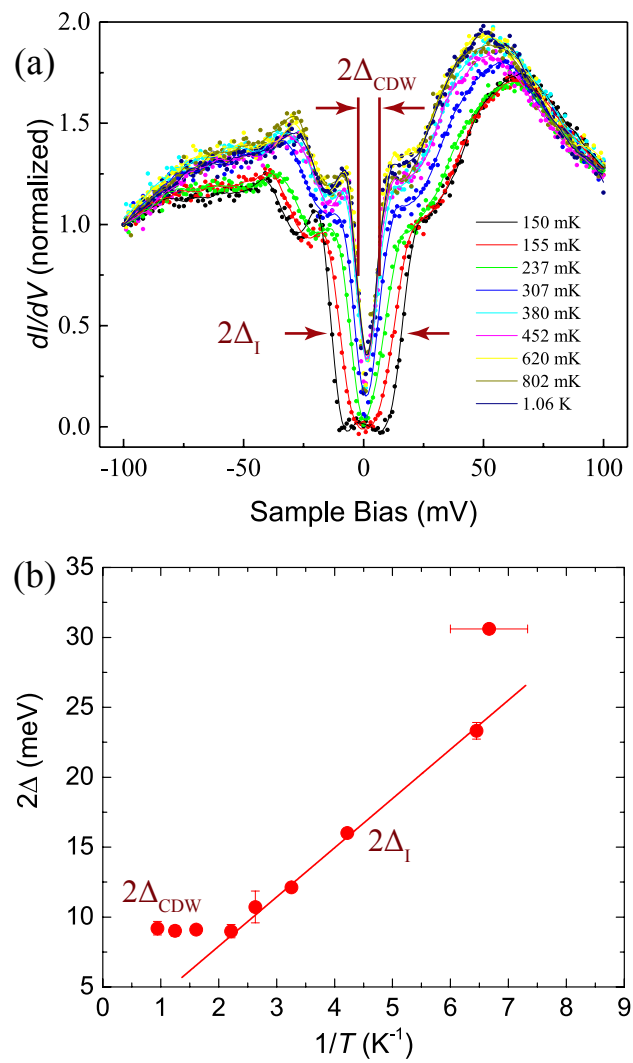

FIGURE 5.

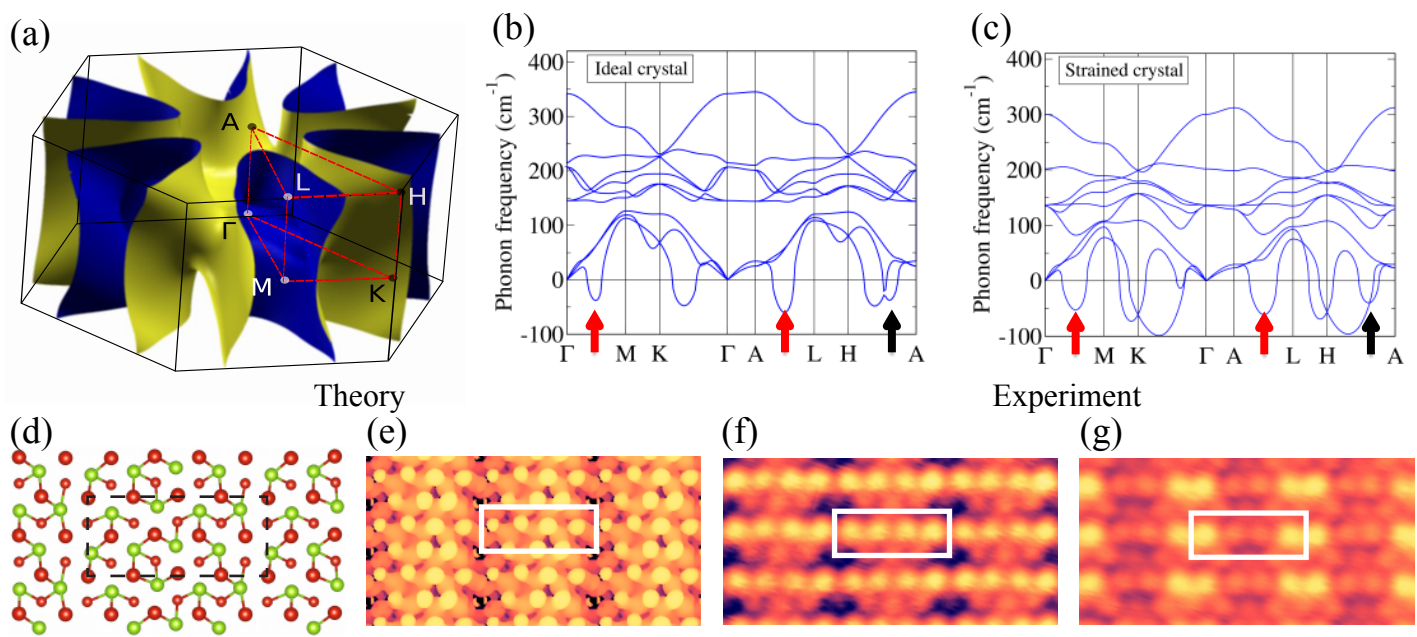

FIGURE 6.